# A RFID-based Campus Context-Aware Notification System

Nazleeni S. Haron, Nur S. Saleem, Mohd H. Hasan, Mazeyanti M. Ariffin and Izzatdin A. Aziz

**Abstract**— This paper presents the design and development of a context-aware notification system for university students using RFID technology. This system is leveraging on the student's matrix card as the RFID tag (sensor), RFID reader and server as the processors and screen monitor at the various locations in the campus as the actuator of the output. This system aims to deliver urgent notifications to the intended students immediately at their respective locations. In addition, the system is also able to display personalized information based on the students' preferences and current location when accessing the system. The background of the study, the design approaches for this system and the preliminary evaluation of the prototype are presented in this paper. The evaluation results have indicated that the the proposed system is useful and easy to use.

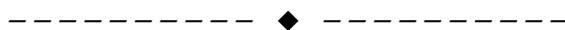

## 1 INTRODUCTION

CONTEXT –AWARE is a term that describes the ability of an application to provide personalized sevice based on the detected context. A system is context-aware if it can extract, interpret and use context information and adapt its functionality to the current context of use[1] .Context can be defined as the information that characterizes the situation of an entity [2]. An entity is a person, place, or object that is considered relevant to the interaction between a user and an application, including the user and applications themselves [2]. The four types of primary context identified in [3] are location, identity, time and activity.

With the advent of context-aware computing, it enables the services to be discovered and performed autonomusly on behalf of the user. As such, many researchers have embedded this technology in their proposed applications [4-9] and the extensive list of developed context-aware applications can be found in [10]. However, despite the popularity of this technology and the advantages it offer, only one out of the many applications has deployed this technology for notification system in a university environment [10].

Therefore, this paper proposes a RFID-based Campus Context-Aware Notification System (R-CCANS) that implements context-aware concept for a notification system in a campus environment using RFID technology. In this research we are trying to prove that the context-aware concept can be applied in a university environment by leveraging on RFID technology.

Our work is motivated by the observation on the existing campus notification channels, E-learning and bulletin board. E-learning is an online learning system that is

————————————————

- *N.S.Haron, N.S. Saleem, M. H. Hasan and M. M. Ariffin are with Department of Computer and Information Sciences, Universiti Teknologi PE-TRONAS, 31750 Tronoh, Perak, MY.*
- *I.A.Aziz is with Deakin University, AUS.*

normally used by lecturers to post academic-related notifications due to its extensive coverage. The problem that exists in E-learning is that, it cannot cater for real-time notification. The students have to log on to the internet to get notified. For instance, the lecturer posted last minute announcement saying that that the class is cancelled. He posted at 8.55 am but the class starts at 9 am. The students who are already waiting in the class will not be notified. Another problem in E-learning is it has a limitation in term of the usage. It cannot be accessed when there is no internet service available.

Another announcement medium was using bulletin board by campus staff for announcing campus events. By using this method, notification will hardly reach the intended and interested audience due to its limited coverage.

Life as a typical students are so hectic and busy. Sometimes, students are not aware of what is happening around them. Therefore, it is vital to have a system that can convey notification to students so that they know what is happening around them. Other than that, by having a notification system it also can help to [11]:
- Increase the likelihood of reaching students immediately with important information
- Improve response times from the students

It has been proven that users tend to ignore screen display notifications if they are too used to it or if they feel that those notifications are not relevant to them. Therefore by providing personalized notifications at the screen, it will intrigue the user to see the contents displayed on the screen that might contain pertinent information to the user. As a consequence, user can make a decision and execute actions accordingly.

R-CCANS is leveraging on RFID technology to ensure the notification reached the intended audience in timely manner (before the preset deadline). To illustrate what motivates the needs for R-CCANS and how R-CCANS works, consider two scenarios as follow:

Jen is an undergraduate student and has a replacement class at 8 pm tonight. It was 5 pm in the evening and Jen was with her colleagues at the café. One of her colleagues, Fay was asking for her help with their soon-to-due class project tonight. Jen was about to answer no when she decided to check the class status at the nearby R-CCANS display. One hour before, Prof. Layne has decided to postpone the class due to unforeseen circumstances. She has posted a notification via R-CCANS website. Jen passed by a reader and a notification from Prof. Layne was displayed. Jen then informed Fay on her availability tonight. As a result, Jen does not have to waste her time going to the cancelled lecture and yet she can still do a fruitful activity tonight.

John is the university sports officer who just received an invitation letter for a friendly soccer match in 2 days time. Due to the urgent needs, he decided to do the audition on that very same day. He publishes the announcement via R-CCANS to those who happened to be in near or at sport complex only. Farris is a graduate student, an avid fan of soccer, really good at playing the game and always linger around at the sports complex to use the facility. He happened to pass by an R-CCANS display and announcement made by John was then shown on the screen. Without hesitation, Farris made his way to the audition venue.

Both of these scenarios depicts that personalized notifications can be published to subset of students based on their registered course or based on their registered preferences and locations. R-CCANS deploys the matrix card as the RFID tag (input to the system), RFID reader and server as the processor and screen display as the actuator.

This paper is structured as follows. We introduce some background concepts of context-awareness and context-aware categories. We describe the RFID technology that we are using. We then present the design of the R-CCANS followed by the prototype development and evaluation. Finally, we draw our conclusions.

## 2 CONTEXT-AWARE COMPUTING

### 2.1 Context-Aware Concept

Context-aware refers to the ability of computing devices or applications to detect and respond and interact in users' environment [12-16]. Other than that, it is also an application that dynamically changes or adapts their behavior based on the context of the application and the user [2].

Context can be classified into four types (abbreviated as TILE) [3]. T is for *Time* context and is usable when the user enters a service zone and if the service is available at that time, then the service is working for the user. I represents *Identity* context that indicates with whom the system is communicating. L stands for *Location* which representing the location of the user or the location of an identity that the user is interested in. Lastly, E is for *Entity* concept that portrays information of any things that the user may be currently using [17].

In addition, a system is said to be context-aware, if it uses context to provide relevant information and / or services to the user, where relevancy depends on the user's task. To develop a good context-aware application, it must be able to answer all the questions below:
- Who is the User?
- Where is the User?
- What is he/she doing?
- What is his requirement?

As a summary, context-aware computing involves application development that allows for collection of context and dynamic program behavior dictated by the knowledge it gathers from the environment [3].

### 2.2 Context-Aware Category

Three attempts have been made in developing taxonomy for context-aware categorization [2, 13, 18]. However, there are some overlapping among these taxonomies and we have chosen to further describe taxonomy proposed by Schilit et al in [17].

TABLE 1
CONTEXT-AWARE CATEGORIES

|  | manual | automatic |
|---|---|---|
| **information** | Proximate selection Contextual information | Automatic contextual reconfiguration |
| **command** | Contextual commands | Context-triggered actions |

According to [18], context-aware applications can be divided into four categories and these categories are the product of two points along two orthogonal dimensions as shown in Table 1. The dimensions divide the categories into whether the task at hand is getting information or doing a command and whether it is affected manually or automatically to the user.

Futher explanation of each category is as follows:.

*Proximate Selection*

In this category, the applications retrieve the information for the user manually on where the located-objects (physical or virtual) resided based on the current location of the user or other specified context. In this case, the application still requires manual intervention from the user and will not choose any specific object automatically. Examples include the application able to display available nearby computer input and output devices such printers, displays, speakers for the users to choose from. Location information can be used to weight the choices of choosing the objects [18].

*Automatic Contextual Reconfiguration*

Applications that are developed under this category actually must be able to retrieve information automatically from the available contexts. They must also able to recon-figure the system to bind to new resource based on new context. Reconfiguration in this case refers to adding new components, removing existing components or altering the connection between components in the system [18]. Reconfiguration could be based on location, activity or the type of user using the applications. For example, when a set of users are having a meeting in a room, a workspace in form of virtual whiteboard is provided. The user can use the whiteboard to share their thoughts and ideas. The moment they enter the room, their mobile host will be reconfigured and binded to the room's virtual whiteboard. And if they were to move to other room for other kind of activity, another resource available in the room will be binded to their mobile host for different usage.

*Contextual Information and Commands*

Contextual Information and Commands makes use the fact that people's actions can be predicted by their situation. In this category, applications offer information and command based on the context in which they were issued. The contextual command can be found in two modes. Firstly, the appearance of the command itself may change depending on context of use [18]. For example, when the user is in the library, a command for invoking card catalogue appears on screen whereas it is normally hidden. Secondly, a command may appear the same but present different results [18]. For example, the same migrate command button appears on the location browser interface while moving from room to room. However, it will invoke to different hosts depending on the location it was invoked.

*Context-Triggered Actions*

Context-triggered actions use simple IF-THEN rules to specify how context-aware systems should adapt. The category is similar to contextual information and commands only except that, it will invoke according to the specified rules [18]. It will invoke automatically when the right combination of contexts exists. For example, monitors the "washing-machine" badge which is attached to the washing machine and will play a sound once it finished washing.

## 3 RFID TECHNOLOGY

RFID is a dedicated short range communication (DSRC) technology. This type of technology has been widely used around the world. Its uses radio waves to automatically identify people or objects. It is a form of automatic identification technology that refers to data capture and storage with no or very minimal human intervention [19]. All RFID data can be read through the human body, clothing and non-metallic materials. The most important aspect of RFID is that it does not require contact or line of sight for communication.

Basic RFID consists of three components [9]. They are:
- An antenna or coil
- A transceiver (with decoder)
- A transponder (RF tag) electronically programmed with unique information

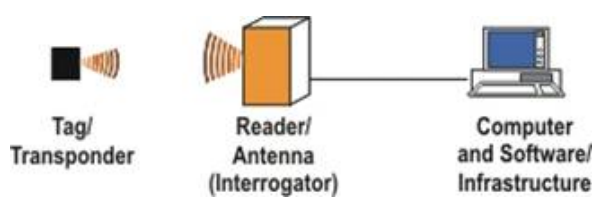

Fig. 1. RFID Infrastructures

Fig. 1 shows the RFID infrastructures. The antenna emits radio signals to activate the tag and at the same time read and write data to it. Antennas are the middle-man between the tag and the transceiver. It controls the system's data acquisition and communication [19]. The reader emits radio waves in ranges of anywhere from one inch to 100 feet or more, depending upon its power output and the radio frequency used. When an RFID tag passes through the electromagnetic zone, it detects the reader's activation signal. The reader decodes the data encoded in the tag's integrated circuit (silicon chip) and the data is passed to the host computer for processing [20].

RFID systems use many different frequencies depending on the type of applications. Available frequencies include low frequency (~ 125 KHz), high-frequency (13.56 MHz) and ultra-high-frequency or UHF (860-960 MHz) while microwave (2.45 GHz) [19].

There are two general categories of RFID systems, active and passive. As for active RFID, it uses active tags that carry their own power source in the chip to generate the radio waves that can be read by a reader from larger distances [19]. In this system, the broadcast waves from the tag can be read over a range to 60 to 300 feet. However, despite this advantage, it has trade off with increased size, cost and reduced operational life. Active RFID systems are normally used for tracking objects that need large range such as cargo container or warehouse tracking. As such the vehicle does not need to slow down or the items that do not being close to the reader.

Passive RFID system is using passive tag of which the tags do not transmit radio waves but communicate by reflecting back the waves they receive from the reader. The tags do not have its own battery and power its circuit using the electromagnetic signal generated by the reader. They are very cheap and small yet it has limitation in terms of reading range (up to twenty feet only). However, this kind of technology is suitable for pet tagging, identification cards or retail merchandise [19].

In this project, we have chosen passive RFID system as the technology for conveying notification system since it requires onetime cost (during the initial purchase and implementation) and it is relatively cheap unlike SMS and active system. This technology is very useful to large batch of target recipients and the best method to avoid spam of notifications. Since it has small reading range, it can provide better personalized notification to the user.

# 4 SYSTEM DESIGN

## 4.1 Context Model

We have chosen three out of four contexts outlined in [3]. Firstly, *Time* context is chosen to represent the timestamp when the students use the system. The obtained value will be used to evaluate and compare with the expiry date of the notification posted by the lecturer and other staff. *Identity* context refers to the student who is going to receive the notification. This context will help to empower the personalization feature. *Location* context is the current nearby building or venue of the student while accessing the system.

We adopt Abowd's dimensions [3] and design approach in [17] in modeling the context. The model of our context is as follows:

Context::==Time_context+Identity_context+Location_context

  Time_context::=Timestamp+Expiry;

    Expiry::=Date+AM/PM+Hour+Minute;

  Identity_context::=Personal_profile;

    Personal_profile::=Tag_id+Course_id+Preferences;

      Preferences::=['Book'|'Class'|'Sports'|'Events'|'misc'];

  Location_Context::=Building_name+Venue_name;

## 4.2 Context-aware Category

We chose the Context-Triggered Actions category [18] since we want the system to perform the required action autonomously based on presented and detected context. The inferring mechanism is using IF-THEN rules and represented as follows:

```
IF<context1>
AND<context2>
….
AND<context(n)>
THEN<display_notification(i)>
```

For example, consider a scenario of which current context is as follows:

Time = '5 pm'
Identity = '1038'
Location = 'Sports Complex'
Preferences = 'Sports'

Then the rules will be presented as follows:

```
IF  hour = 5
AND ampm = 'pm'
AND location = 'sports_complex'
AND preferences = 'sports'
THEN display_notification = "inter-varsity football
league is on now"
```

## 4.3 System Architecture

Fig. 2 shows the proposed system architecture for the R-CCANS. When the student passes by any RFID reader, the reader will detect the presence of a tag (embedded in the student's matrix card). In this system, we will be using a passive tag. Each student's matrix card is assigned a unique tag ID and each reader is assigned a unique location ID. The server will find the respective information in the database based on the detected tag ID and location ID. Once the server found the ID and the matching notification, it will then display the notification on the screen monitor nearby to the student. Therefore, as long as the students are within 1 inch to 20 feet, the reader will be able to detect the students and the personalized notification can be displayed on the screen. We have chosen the screen as the display unit instead of the user personal devices (PDA, mobile phones or laptop) due to implemetation limitation. This is because the implementations and API features of various devices are different. Therefore, in order to solve that we have to develop different type of client interface. In addition, it will be hard to manage in case the client interface needs to be updated.

The personalized information was derived based on preferences and user profiles acquired manually from the user. The users have to perform manual registration with the system and their profiles and preferences will be stored in information repository for future use.

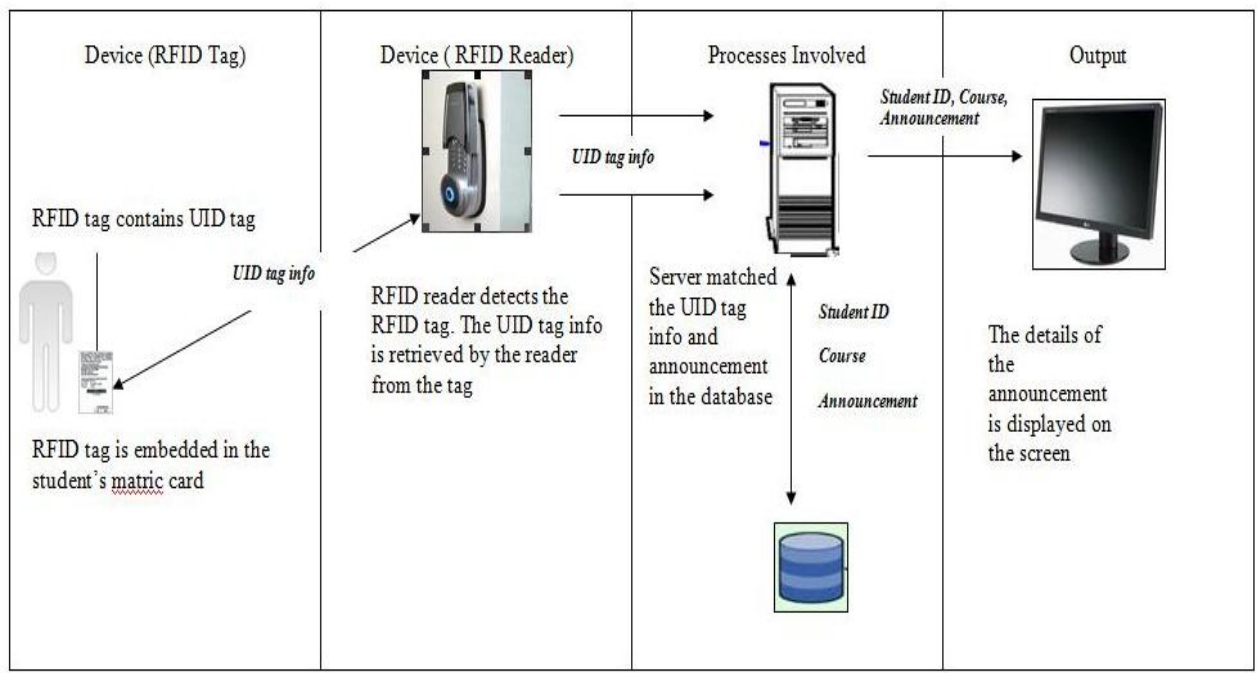

Fig. 2. System Architecture

## 5 PROTOTYPE DEVELOPMENT

### 5.1 Hardware

For prototyping, we have used the UPM Raflatac Dog-Bone HF RFID as the passive tag for the proposed system. We used the Passive HF RFID Reader of Pocket Reader Juno7 as the reader to extract the information stored in the passive RFID tags. We also used the same RFID reader as the display component to display the notification for our prototype system.

### 5.2 Software

We have developed the prototype using Microsoft Visual Studio 2008 development kit and specifically using VB.Net and C# as the programming languages. These languages are chosen due to the simplicity in coding for interfacing with the RFID hardware. In addition, we chose Microsoft Access for storing the information (database) and JUNO_900W_C# as the middleware for RFID devices.

### 5.3 User Interface

There are two main types of user interface of R-CCANS. One is for the users who send the notification (eg. lecturer or other university staff) and the other one is for the recipient of the notification (student).

*Post Notification User Interface*

Fig.3 shows the posting announcement page. This is where the lecturer or other staffs posts the announcement. The created date is automatically produced to avoid any error in determining the notification creation date. The sender's name will be automatically inserted based on the login profile. The notification expiry date is to be determined by the lecturer.

Prior to the expiry date, the contents of the notification will be displayed to the intended students. Once the notification reaches the expiration date, it will no longer be notified to the students. The lecturer/staff can choose either specific student or a batch of student from a specific course as the target recipient for the notification. The list of students and courses are available for retrieval from the database. The related location of the notification will be inserted in the next page together with extended details of the event.

The lecturer/staff can also search for announcement made earlier. The search can be made by specifying either by who posted the notification, by the notification creation date or by the notification title.

*Notification Display User Interface*

The user interface is as shown in Fig. 4. The student will be presented with the latest personalized notifications. Student can choose to read any notification displayed. Student can also delete notification and mark notification as unread in case they want the notification to appear as unread. The close button is to close the announcement page.

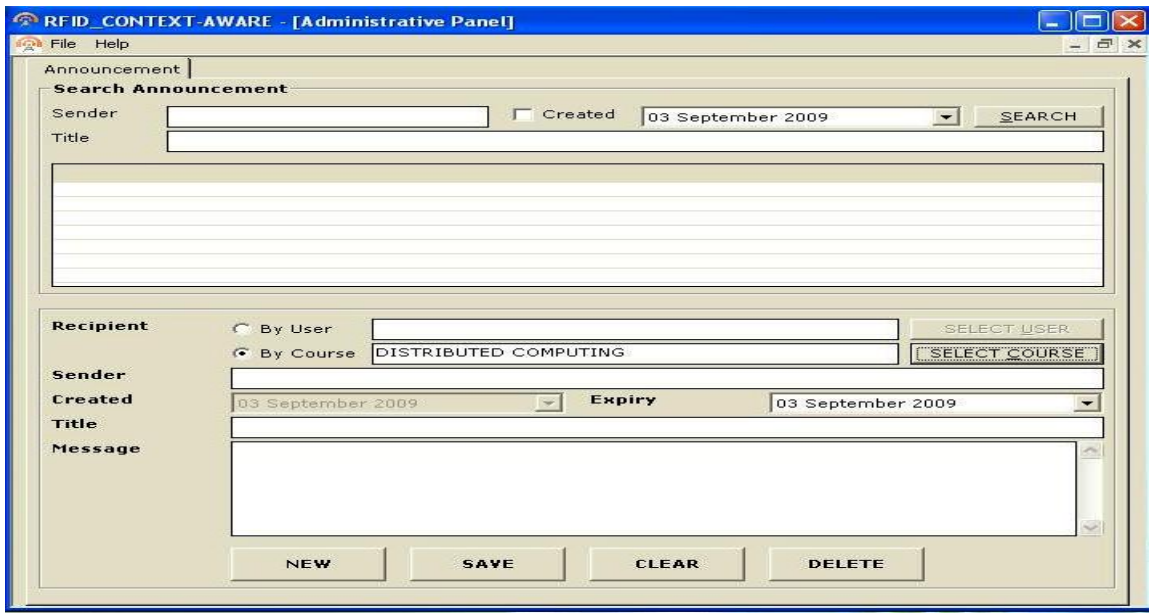

Fig. 3. Post Notification User Interface (for lecturers or staff)

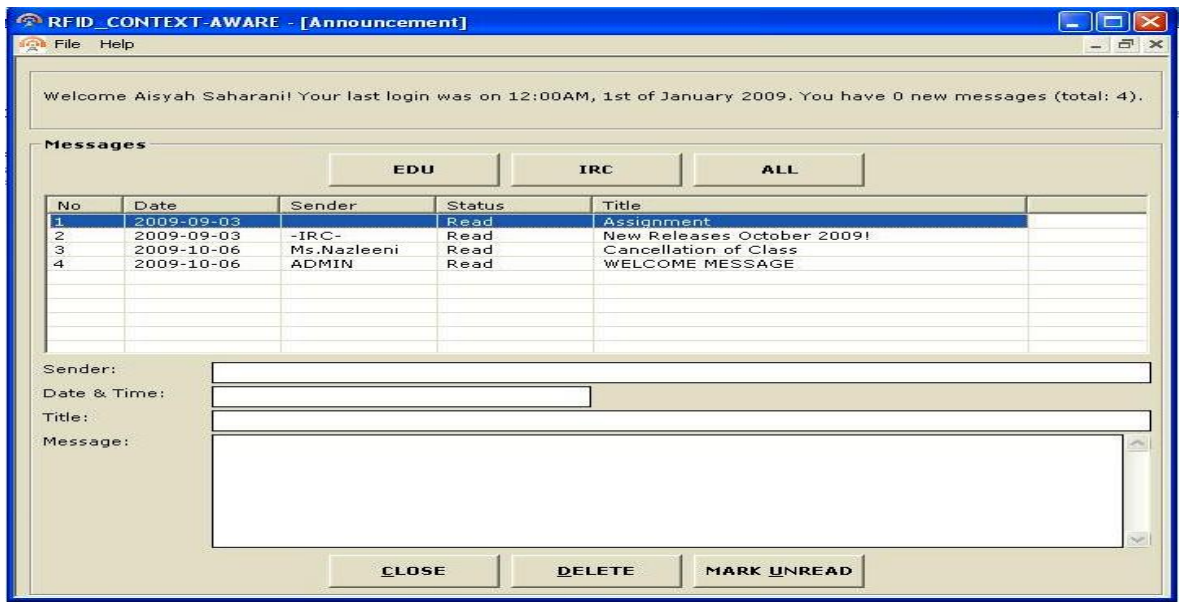

Fig. 4. Display Notification User Interface (for student)

## 6 PROTOTYPE EVALUATION

We have a conducted an evaluation testing in order to solicit the usefulness and ease of use of the proposed system. The respondents were from various backgrounds but chosen from two groups of university members: lecturers and undergraduate students. To evaluate the system, we have administered a questionnaire that was designed based on Technology Acceptance Model (TAM) [21]. The respondents have to fill up the questionnaire once they have completed experiencing with the system. 20 statements/items are prepared altogether for both categories and for both types of respondents. They have to rank between one(1) to seven (7) for each statement in the questionnaire given. one (1) means strongly agree while seven (7) represents strongly disagree.

Fig. 5 presents the result on usefulness and ease of use evaluation collected from lecturers. The lecturers collectively ranked usefulness at 2.3 and this demonstrates that system is useful to them. As for ease of use, in average the lecturers ranked it at 2.5 and this reveals that the lectures find it easy to use the proposed system.

Fig. 6 shows the students' point of view on both usefulness and ease of use. They have ranked usefulness at 2.4 and ease of use at 1.8. As such, it can be concluded that the students also find the system both useful and easy to use.

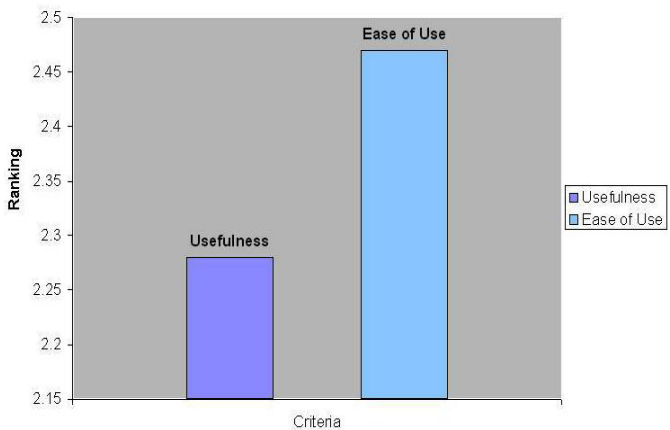

Fig. 4. Results on Usefulness and Ease of Use collected from lecturers

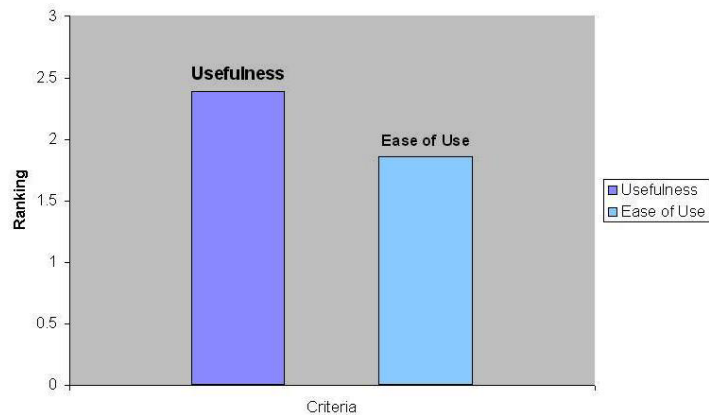

Fig. 5. Results on Usefulness and Ease of Use collected from students

## 7 CONCLUSION

In this paper, we propose a design for a RFID-based Campus Context-Aware Notification System (R-CCANS) which aims to deliver notification to the intended recipients in timely manner leveraging on RFID technology. This system uses contextual attributes such as time, location and identity of the user in inferring the right notification to display. R-CCANS is context-triggered actions system where it acts autonomously when the context is detected (via tag ID and location ID) and processed. The notification given is also derived from user's personal preferences which have been acquired by manual registration. The preliminary evaluation on the prototype conducted on two types of user has indicated that the system is useful and easy to use. However, the prototype is still in progress and we have envisioned to add more features in the future R-CCANS. Among the planned functions are the ability for the student to post notification as well as for the message to be able to be transferred to user's mobile device via Bluetooth or Infrared technologies.